# Estimation of the dome seeing from results of the optics quality tests with Shack-Hartman wavefront sensor.

S. A. Potanin[a]

[a]Sternberg Astronomical Institute Moscow State University

## ABSTRACT

The Shack-Hartman wavefront sensor designed for final acceptance of 2.5m SAI telescope allows to measure the shape of the wavefront on exit pupil of the telescope using a bright star. The reference laser source on wavelength 532 nm is used for measurements. During testing the device at different telescopes it was found out what it's probable to estimate some properties of air streams in the dome. Such estimations have been made for three domes: 1.5 m AZT-22 and 1.0 m Zeiss telescope of Maidanak Observatory, Uzbekistan and for 2.6 m telescope ZTSh of Crimean observatory, Ukraine. The following results were obtained for the slowest streams in domes: $\beta=0.34"$ for AZT-22, $\beta=0.67"$ for Zeiss-1000, and $\beta=0.69"$ for ZTSh.

## 1. INTRODUCTION

The work on creation of Shack-Hartman wavefront sensor for testing of optics quality of the installed telescopes has been finished in 2009(fig.1). The device is intended for testing of optics quality of 2.5 meter telescope of SAI Moscow State University. During device testing a number of factors disturbing measurements of this kind has been revealed. Slow streams of air under the telescope dome is one these factors. For the effective averaging of them it was necessary to do hundreds and thousands of separate measurements[1]. It had given the opportunity to carry out further estimations of some characteristics of these streams. It is necessary to take into consideration the reasons of movement of air in a dome. The most principal reasons are the streams from warm subjects (including from the primary mirror of the telescope) and streams produced by external wind.

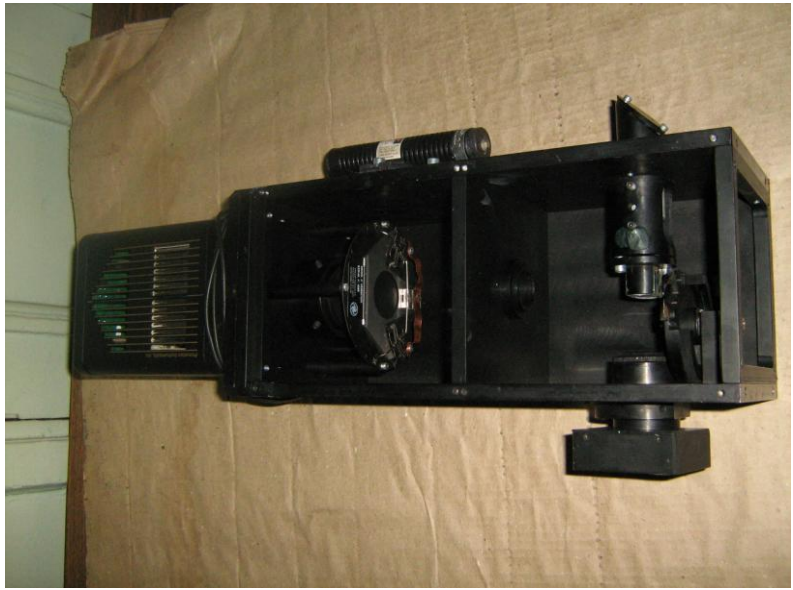

Figure 1. Photo of Shack-Hartman wavefront sensor. The top cover is open.

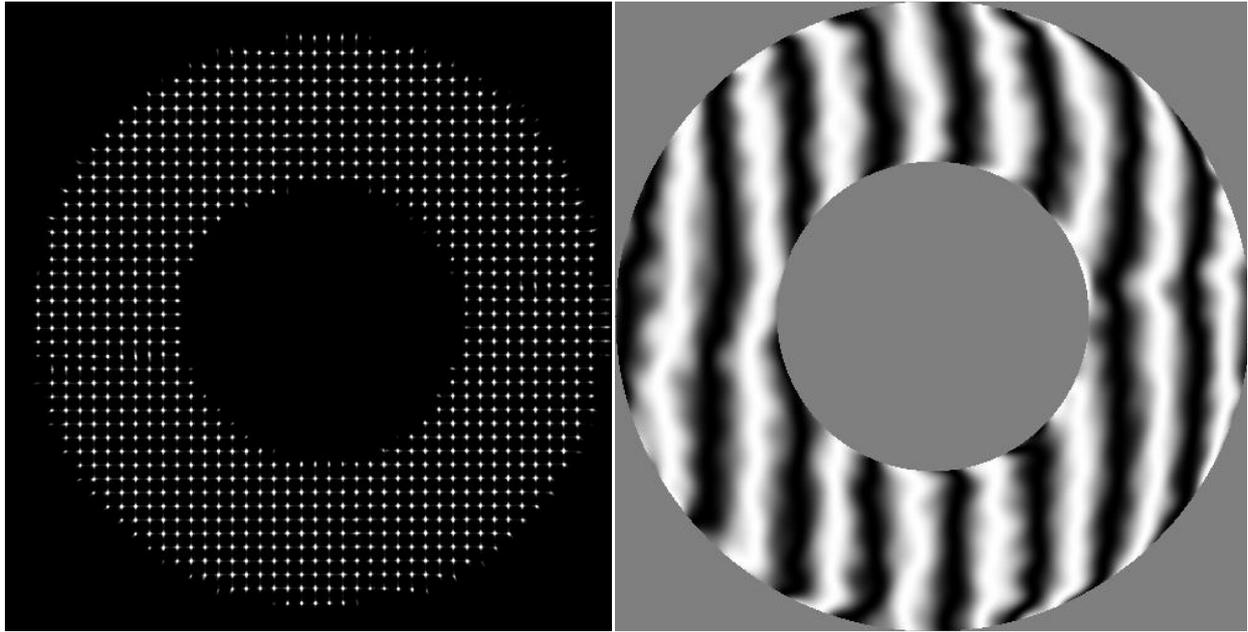

Figure 2. The hartmanogram and the total synthetic interferogram (at 532 nm system one pass) of 1.5m telescope AZT-22 of the Maidanak observatory. Optical quality: wavefront RMS=37nm.

Optical tests with this device was carried out for 6 telescopes from 0.7 to 2.6 m. In all cases it turned out to be possible to receive an estimation of dome seeing for slow (quasistationary) streams.

## 2. DOME SEEING ESTIMATION TECHNIQUE

The technique of measurement of dome-seeing is similar to the technique applied in devices DIMM[2,4] and S-DIMM[3]. The results of running sensor on a telescope are the measurements of local tilts of wavefront in subapertures. Thus we have a set of the fixed bases and for each base we can calculate a dispersion of differences of tilts in subapertures on the base ends. Than these dispersions were plotted against the length of bases. This function is actually the structure functions of second derivative of phase and can be compared with the standard Kolmogorov model of free atmosphere what we done for 3 telescopes.

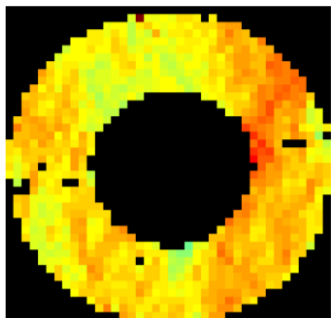

Figure. 3. Individual image with tilts in X direction.

# 3. RESULTS

The photo of towers and structure functions for 3 telescopes are shown below. Measurements were carried out in 2008-2009.

## 3.1. AZT-22 Maidanak Uzbekistan

Telescope diameter: 1500 mm F/7.7
Effective subaperture: 37.5 mm
Exposure time: 100 ms
Time between exposures: 5 s

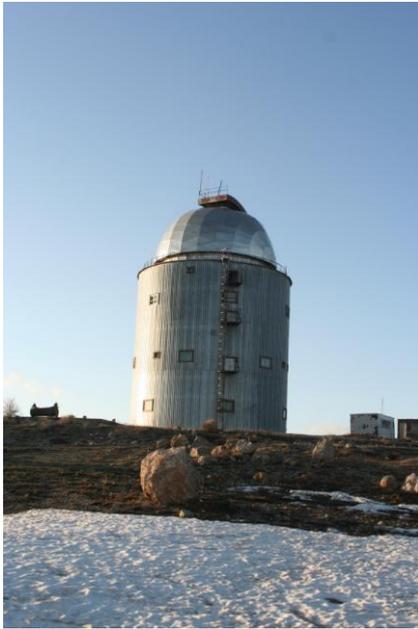 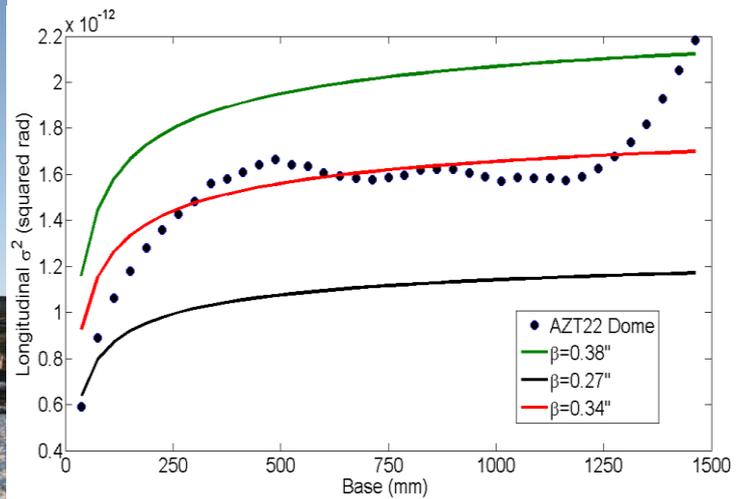

Figure 4. AZT-22 tower (left). Longitudinal tilt variance as function of subapertures separation and theoretical curves (standard model for free atmosphere) for different values of the seeing. (AZT-22 Maidanak)

AZT-22 dome seeing estimate: 0.34 arcsec.

## 3.2. Zeiss-1000 Maidanak Uzbekistan

Telescope diameter: 1000 mm F/10
Effective subaperture: 41.7 mm
Exposure time: 300 ms
Time between exposures: 6.5 s

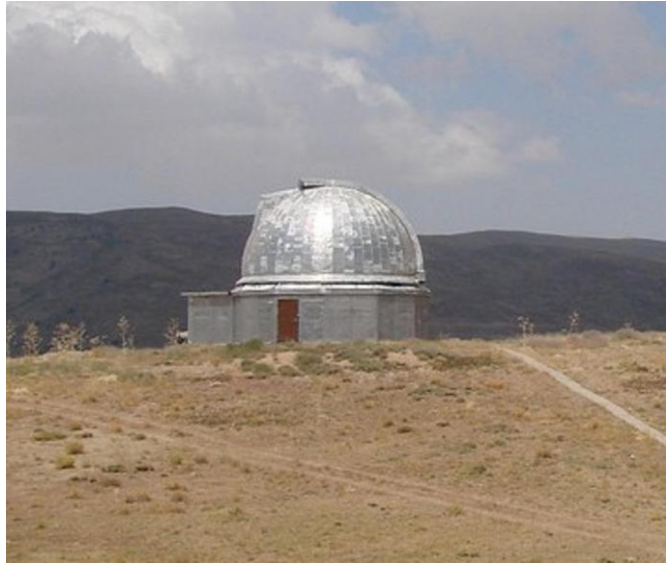

Figure 5. Zeiss-1000 tower

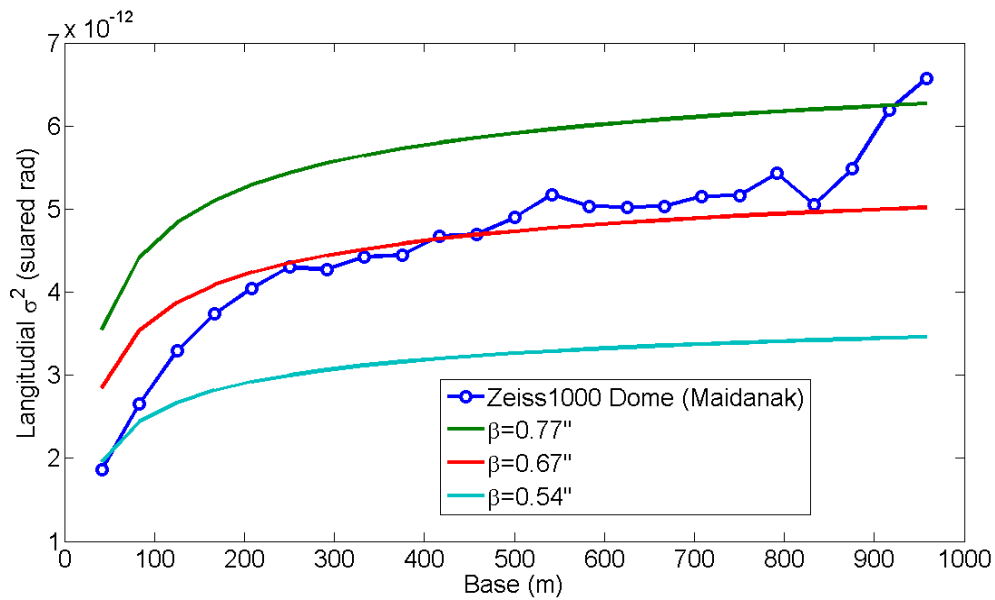

Figure 6. Longitudinal tilt variance as function of subapertures separation and theoretical curves (standard model for free atmosphere) for different values of the seeing. (Zeiss-1000 Maidanak)

Zeiss-1000 dome seeing estimate: 0.67 arcsec.

### 3.3. ZTSh Crimean observatory Ukraine

Telescope diameter: 2600 mm F/16
Effective subaperture: 130 mm
Exposure time: 100 ms
Time between exposures: 7.7 s

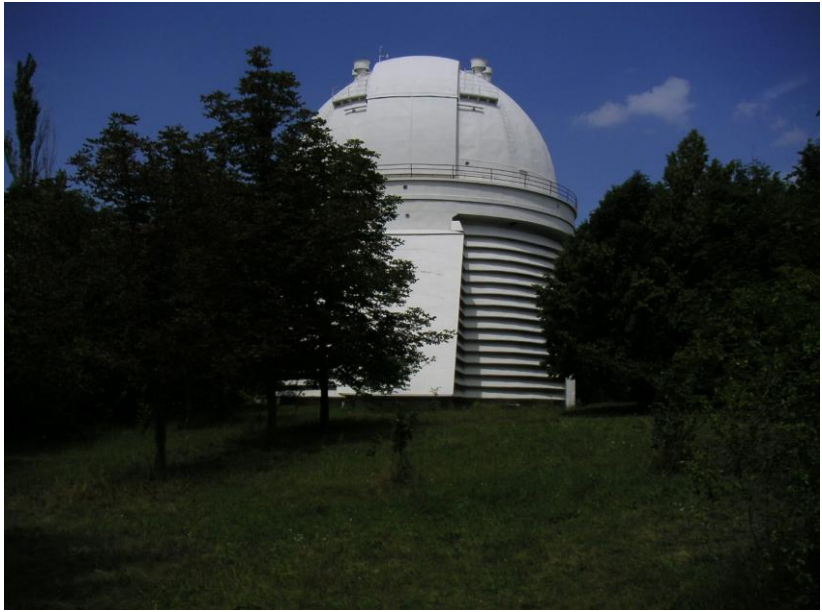

Figure 7. ZTSh tower

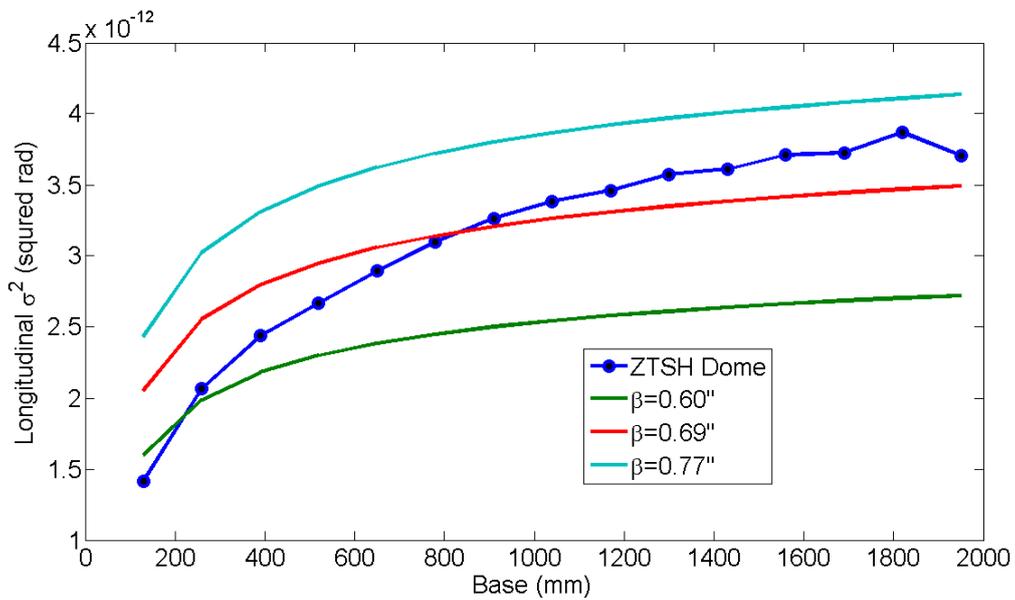

Figure 8. Longitudinal tilt variance as function of subapertures separation and theoretical curves (standard model for free atmosphere) for different values of the seeing. (ZTSh Crimea)

ZTSh dome seeing estimate: 0.69 arcsec.

## 4. CONCLUSIONS

So, by means of the wave front sensor it was possible to make estimations of seeing for slow air currents under the dome. Strictly speaking these data are also influenced by slow streams in atmosphere outside the dome on a sight direction. But the presence of correlations on scale of the size of apertures (see fig. 5 and 6) indicates that this influence is not very large and most part of of observable air currents occur in a dome.

Domes of ZTSh and Zeiss-1000 demonstrate much worse situation than for AZT-22 (0.34"). For ZTSH estimated dome seeing is 0.69", and for Zeiss-1000 - 0.67". It can be explained that the Zeiss1000 dome is low and the ZTSH tower is surrounded by high trees.